\documentclass[twocolumn,aps]{revtex4}
\usepackage{graphicx}
\usepackage{dcolumn}
\usepackage{amsmath}
\usepackage[latin1]{inputenc}
\newcommand{\ekp}{eight-band \mbox{${\bf k}\!\cdot\!{\bf p}$}}

\begin{document}

\title{Size-dependent fine-structure splitting in self-organized InAs/GaAs quantum dots}

\author{R. Seguin\footnote{Email: seguin@sol.physik.tu-berlin.de}, A. Schliwa, S. Rodt, K. Pötschke, U. W. Pohl, and D. Bimberg}

\affiliation{Institut f\"ur Festk\"orperphysik, Technische
Universit\"at Berlin, Hardenbergstrasse 36, 10623 Berlin, Germany}

\begin{abstract}

A systematic variation of the exciton fine-structure splitting with quantum dot size in single InAs/GaAs quantum dots grown by metal-organic chemical vapor deposition is observed. The splitting increases from -80 to as much as \mbox{520 $\mu$eV} with quantum dot size. A change of sign is reported for small quantum dots. Model calculations within the framework of \ekp\ theory and the configuration interaction method were performed. Different sources for the fine-structure splitting are discussed, and piezoelectricity is pinpointed as the only effect reproducing the observed trend.

\end{abstract}

\maketitle

The exchange interaction of electron-hole pairs (excitons) in semiconductor quantum dots (QDs) has been subject of a lively debate in recent years \cite{bayer02, kowalik05, lenihan02, hoegele04, tartakovskii04b, langbein04, finley02, bester03}. 
In such strongly confined systems it is supposed to be enhanced with respect to the bulk case due to the close proximity of electrons and holes. However, the influence of the exact geometry of the confining potential on the exchange interaction still needs to be clarified. A detailed understanding of the resulting exciton fine structure in quantum dots is of fundamental interest and of largest importance for potential applications of QDs in single-photon emitters and entangled two-photon sources for quantum cryptography \cite{benson00}.

The total angular momentum $M$ of heavy-hole excitons (X) in QDs is composed of the electron spin ($s=\pm \frac{1}{2}$) and the heavy hole angular momentum ($j = \pm \frac{3}{2}$), consequently producing four degenerate exciton states frequently denoted as \emph{dark} ($M=\pm 2$) and \emph{bright} ($M=\pm 1$) states indicating whether they couple to the photon field or not. Independent of the given confinement symmetry electron-hole exchange interaction causes a \emph{dark-bright} splitting. Furthermore it mixes the \emph{dark} states lifting their degeneracy and forming a \emph{dark} doublet $(|2\rangle \pm |-2\rangle)$. Likewise, additional lowering of the confinement symmetry to $C_{2v}$ or lower mixes the \emph{bright} states producing a nondegenerate \emph{bright} doublet $(|1\rangle \pm |-1\rangle)$.

\begin{figure}[ht]\centering
\begin{center}
\resizebox*{0.7\columnwidth}{!}{\includegraphics[angle=0]{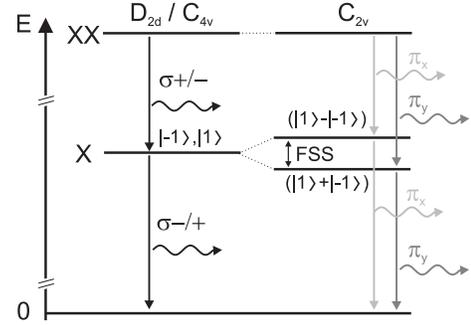}}
\end{center}
\caption{Energy scheme illustrating the effect of the exchange interaction on the exciton (X) bright states. XX denotes the biexciton ground state. $|M\rangle$ is the exciton total angular momentum, $\sigma\!\!+\!\!/\!-$ denotes circularly polarized light, and $\pi_{x/y}$ linearly polarized light with $x=[110]$ and $y=[1\bar{1}0]$. $D_{2d}$, $C_{4v}$, and $C_{2v}$ indicate the confinement potential symmetry.}
\label{eschema}
\end{figure}

While emission lines involving pure states are circularly polarized, the mixed states usually produce lines showing linear polarization along the $[1\bar{1}0]$ and $[110]$ crystal directions, respectively (Fig.\ \ref{eschema}). 
The two bright states are thus directly observable as linearly polarized transitions in luminescence experiments. 
The energetic difference between these lines is called exciton fine-structure splitting (FSS). 

The biexciton (XX) ground state is not split by the exchange interaction, since the net spin of the involved electrons and holes is 0. However, the XX to X decay involves two allowed transitions with the final states being the bright states of the X. Therefore, the FSS is reproduced (yet inverted) in the XX to X decay (Fig.\ \ref{eschema}).

Recently reported experimental values of the FSS in In(Ga)As/(Al)GaAs QDs are listed in Table \ref{literature}. Its magnitude has been shown to be influenced by annealing \cite{tartakovskii04b,langbein04}, and external electric fields \cite{kowalik05}. A systematic analysis of the influence of the geometry of the confining potential and an explanation for the wide range of observed energies however, is still missing.

Although most of the cited experimental references in Table \ref{literature} provide some theoretical background, only Bester et al.\ \cite{bester03} linked the FSS to morphological QD properties in a theoretical work. 
Using an empirical pseudopotential-based approach including the configuration interaction (CI) they analysed In$_{x}$Ga$_{1-x}$As dots of various shapes and composition. The largest FSS they obtained was 30~$\mu$eV for an elongated QD with a lateral aspect ratio of $1.3$. 
They pointed out that even a cirular based QD having a structural $C_{\infty}$ symmetry exhibits a non-zero FSS due to the atomistic asymmetry of the underlying lattice. 
In their work, however, they did not account for the piezoelectricity-induced asymmetry, which plays a major role in our interpretation of the large FSS observed.

{\small
\noindent

\begin{table}

\caption{Examples of experimental values of the exciton fine-structure splitting (FSS) in quantum dots. Column 1 gives the material system and column 3 the experimental setups used ($\mu$PL = microphotoluminescence, FWM = four-wave mixing).}

\begin{ruledtabular}

\begin{tabular}{c p{3mm} c p{4mm} c p{6mm} c}
QDs/matrix&&FSS [$\mu$eV]&&setup&&Ref.\\
\hline
InAs/GaAs&&110-180&&$\mu$Pl&&\cite{bayer02}\\
InAs/GaAs&&$\le$ 140&&$\mu$Pl&&\cite{kowalik05}\\
InAs/GaAs&&40&&FWM&&\cite{lenihan02}\\
InGaAs/GaAs&&30-150&&$\mu$Pl&&\cite{bayer02}\\
InGaAs/GaAs&&10-42&&transmission&&\cite{hoegele04}\\
InGaAs/GaAs&&8-36&&pump \& probe&&\cite{tartakovskii04b}\\
InGaAs/GaAs&&6-96&&FWM&&\cite{langbein04}\\
InAs/AlGaAs&&500-1000&&$\mu$Pl&&\cite{finley02}\\
\end{tabular}

\end{ruledtabular}

\label{literature}

\end{table}
}

Major effects that lead, in principle, to a lowering of the confinement symmetry include (a) structural elongation of the QDs, (b) strain-induced piezoelectric fields \cite{grundmann95}, and (c) interfacial symmetry lowering and its enhancement by  atomistic elasticity \cite{bester03,bester05}. Since the latter effect has been shown to produce FSS values smaller than 10 $\mu$eV \cite{bester03}, we concentrate on the two former effects in our modeling.

We present in this Letter a systematic study of the FSS in single InAs/GaAs QDs by experiment and theory. 

The sample investigated was grown by metal-organic chemical vapor deposition on GaAs(001) substrate. A 300 nm thick GaAs buffer layer followed by a 60 nm Al$_{0.6}$Ga$_{0.4}$As diffusion barrier and 90 nm GaAs were grown. For the QD layer nominally 1.9 monolayers of InAs were deposited followed by a 5 s growth interruption. Subsequently, the QDs were capped with 50 nm of GaAs. Finally, a 20 nm Al$_{0.33}$Ga$_{0.67}$As diffusion barrier and a 10 nm GaAs capping layer were deposited. Antimony was added during the deposition of the InAs layer and the subsequent growth interruption. 

The sample was examined with a JEOL JSM 840 scanning electron microscope equipped with a cathodoluminescence setup \cite{christen91}. It was mounted onto a He flow cryostat which provided temperatures as low as 6 K. The luminescence was dispersed by a 0.3 m monochromator equipped with 1200 or 1000 lines/mm gratings. Above \mbox{1.2 eV} the light was detected with a liquid-nitrogen cooled Si charge-coupled-device camera and below \mbox{1.2 eV} with a liquid-nitrogen cooled InGaAs diode array. At \mbox{1.28 eV} the minimal spectral resolution as given by the setup was $\approx$140 $\mu$eV.

In order to reduce the number of simultaneously detected QDs a metal shadow mask with circular apertures of 100 and 200 nm in diameter was applied onto the sample surface. This technique allows the reproducible investigation of single QDs.

The QD ensemble luminescence peak is centered at 1.18 eV with a FWHM of $\approx$ 150 meV. It displays a pronounced modulation which is due to a shell-like size distribution of the QDs producing a number of different subensembles. The QDs display a large substate splitting suggesting In/Ga interdiffusion to be negligible \cite{heitz05}. In addition, transmission electron microscopy (TEM) images taken from similar samples indicate a truncated pyramidal shape \cite{pohl}. Therefore the exciton energy is directly connected to QD size.

\begin{figure}[ht]\centering
\begin{center}
\resizebox*{0.99\columnwidth}{!}{\includegraphics[angle=0]{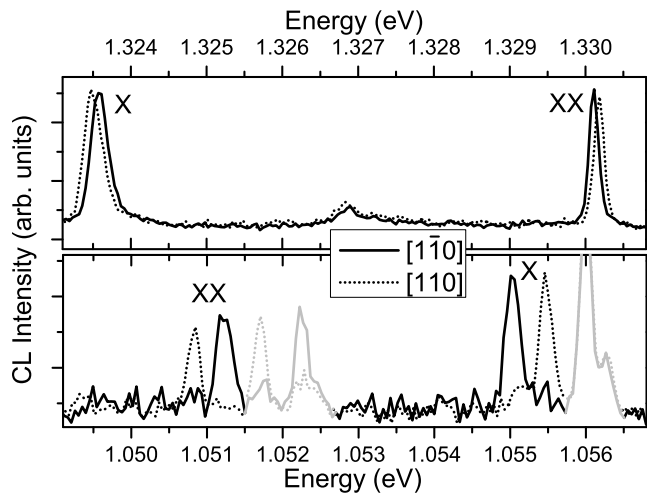}}
\end{center}
\caption{Polarized spectra for two different QDs emitting at high and low energies are shown. The FSS is $-80$ and 420~$\mu$eV, respectively. Gray lines in the lower panel originate from charged excitonic complexes not considered in this Letter. }
\label{spektren}
\end{figure}

When measured through the apertures the luminescence decomposes into single sharp lines. We are able to unambiguously identify spectra of single QDs consisting of up to six lines via their characteristic spectral diffusion pattern \cite{rodt05}. From these lines the exciton (X) and biexciton (XX) transitions can be extracted by performing excitation- and polarization-dependent measurements. Thus the FSS in the corresponding QD can be directly measured. Figure~\ref{spektren} shows examples of polarized X and XX emissions for two different QDs. 

The FSS shown in Fig.\ \ref{spektren} is $-80$ $\mu$eV for the QD emitting at high energies and 422 $\mu$eV for the QD emitting at low energies. The FSS is defined to be positive if the X line at lower energy is polarized along the $[1\bar{1}0]$ crystal direction. The order of the X and XX lines is reversed due to a varying impact of correlation in the QDs as has been established earlier \cite{rodt05}.

\begin{figure}[ht]\centering
\begin{center}
\resizebox*{0.99\columnwidth}{!}{\includegraphics[angle=0]{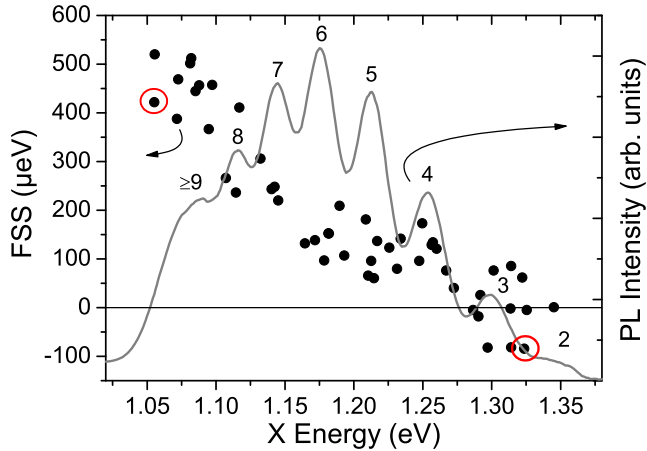}}
\end{center}
\caption{Measured exciton fine-structure splitting (FSS) as a function of exciton recombination energy for 49 QDs. The gray curve depicts the QD ensemble luminescence with numbers denoting QD height in units of InAs monolayers. A clear trend of the FSS is observed. The data points extracted from Fig.~\ref{spektren} are indicated by red (gray) circles.}
\label{data}
\end{figure}

Figure~\ref{data} shows measured FSS values as a function of the exciton recombination energy, i.e., the center between the two X lines. Over the whole energy range of QD luminescence a systematic variation of the FSS within the very same sample is observed. Small QDs that emit at high energy show the smallest FSS, while large QDs that emit at lower energies show the largest FSS. This trend is surprising, since the \emph{dark-bright} splitting has been shown to become larger the smaller the QDs are \cite{bayer02}.

The largest obtained value amounts to 520 $\mu$eV, exceeding all previously reported FSS values in the In(Ga)As/GaAs material system, where the FSS is reported to be always smaller than 180 $\mu$eV (Tab.\ \ref{literature}). Only Finley et al.\ observed in the InAs/AlGaAs system values up to 1 meV \cite{finley02}. They proposed the stronger confinement due to the AlGaAs matrix as a possible reason.

As the FSS is related to a lateral anisotropy of the confining potential, we investigate the two most promising candidates leading to such an effect, namely an elongation of the QD along $[110]$  or $[1\bar{1}0]$ direction and strain-induced local piezoelectric fields. A thorough theoretical investigation requires two essential ingredients: First, precise information about the QD structures involved and, second, a theory that reliably links the QD structure with their electronic and optical properties. 

\textit{Structural assumptions}: In Refs.~\cite{heitz05,pohl} it is demonstrated that the QDs are of pure InAs and have the shape of truncated pyramids with abrupt interfaces. The modulation of the low-intensity-photoluminescence (PL) (see Fig.\ \ref{data}) directly allows to determine the exact height of the QDs belonging to different subensembles. We consider both square-based QDs and, although plan-view TEM data show no indication for a major QD elongation \cite{pohl}, QDs elongated along the $[1\bar{1}0]$ crystal axis. The base lengths of the model QDs are chosen in a way that the exciton spectrum matches our PL and PL-excitation spectra \cite{heitz05} and TEM data \cite{pohl}. The smallest square-based QD has a base length of 10.2 nm and a height of 3 monolayers, the largest one a base length of 15.8 nm and a height of 13 monolayers. Elongated QDs were designed to have the same volume as their square-based counterpart with the same height.

\textit{Method of calculation}: The excitonic states are calculated using the CI method by expanding the exciton Hamiltonian into a basis of antisymmetrized products of single particle wave functions (6 electron and 6 hole wave functions), thus accounting for direct Coulomb interaction, correlation, and exchange \cite{rodt05}. The single particle orbitals are derived from a strain-dependent \ekp\, Hamiltonian accounting for band coupling and band mixing, following our previous work \cite{stier99}. Absorption is calculated by Fermi's golden rule applied to CI states. 

\begin{figure}[floatfix]\centering
\begin{center}
\resizebox*{0.99\columnwidth}{!}{\includegraphics[angle=0]{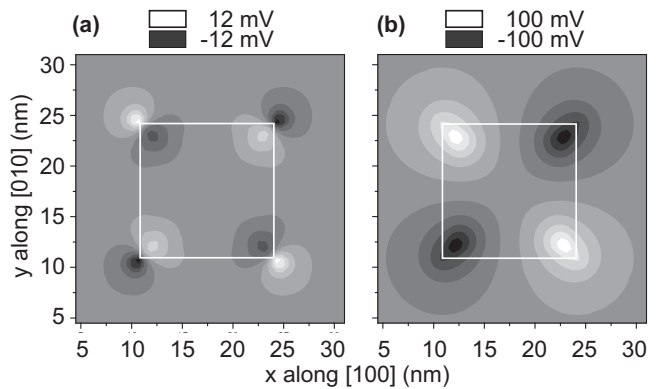}}
\end{center}
\caption{The piezoelectric potential calculated for (a) the InAs bulk and (b) the extrapolated $e_{14}$ value. The potential is shown for a lateral cut near the base of the QD (white square), which is modeled as a truncated pyramid having a base length of 12 nm. } 
\label{piezo}
\end{figure}

To assess the role of piezoelectricity quantitatively, an accurate value of the prime input parameter, the piezoelectric constant $e_{14}$ must be known. However, as Bester et al.\ pointed out recently \cite{bester05}, the standard reference for the InAs piezoelectric constant $e_{14}=0.045$~C/m$^2$ \cite{adachi90} cannot be considered a trustworthy value for strained InAs QDs. More recent measurements \cite{hogg93, sanchez94, chan98, cho03} on strained InGaAs quantum wells grown on GaAs(111) substrate suggest an $e_{14}$ value with a different sign and a much larger magnitude. Because of the apparent strain dependence of $e_{14}$, the exact value for strained InAs QDs is unknown. To encounter this problem we extrapolated the values of the most recent work \cite{cho03} to obtain the value of InAs. Remarkably, the obtained value $e_{14}=-0.385$~C/m$^2$ has a different sign than the `standard' value for bulk InAs as has been noted before by Bester et al.\ \cite{bester05}. Since there is no direct experimental confrimation for this value, quantitative results of the calculations have to be treated with caution (see below). In addition, we point out that the exchange interaction was calculated including only monopole terms. Calculations including dipole terms are under way.

\begin{figure}[floatfix]\centering
\begin{center}
\resizebox*{0.99\columnwidth}{!}{\includegraphics[angle=0]{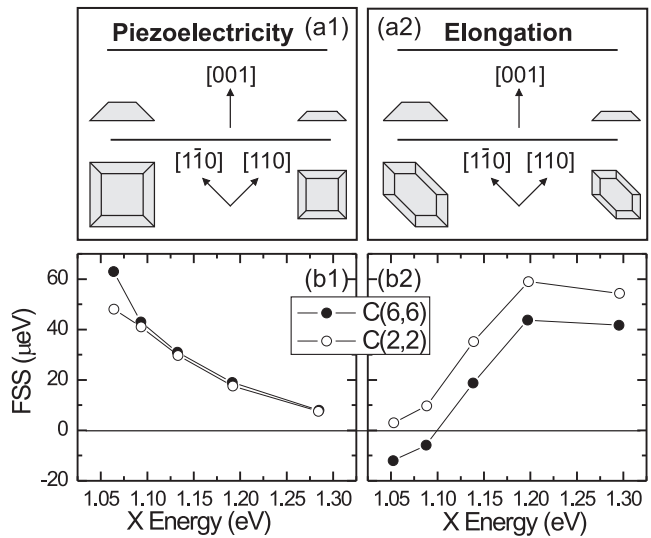}}
\end{center}
\caption{Calculated exciton fine-structure splitting (FSS) values for a series of QDs with a square base including piezoelectricity (b1) and a series of QDs elongated in the $[1\bar{1}0]$ crystal direction in absence of piezoelectricity (b2). C($n$,$n$) denotes the number of electron and hole states included in the CI calculation. The structural assumptions of the modeled series are shown in (a1) and (a2).}
\label{model}
\end{figure}

The impact of the `new' $e_{14}$ value compared to the standard value on the piezoelectric potential is threefold (see Fig.\ \ref{piezo}):
First, and of most importance for the discussion in this Letter, the magnitude of the piezoelectric potential inside the QD is strongly enhanced by a factor of around ten. Second, the field orientation inside the QD is rotated by 90° and third, the charge distribution near the QD corners change from dipole-type to monopol-type.

Although both piezoelectricity and elongation introduce an anisotropy of the confining potential, their impact on the orientation of the participating electron and hole wave functions is fundamentally different: Piezoelectricity forces electron and hole groundstate wave functions to point in orthogonal directions, whereas the groundstate wave functions located in an elongated QD accommodate their orientation to that of the QD, thus being parallel to each other.

To distinguish the impact of elongation and piezoelectricity, we calculated the FSS for a series of square-based QDs \textit{including} the effect of piezoelectricity [Fig.\ \ref{model} (a1)] and a complementary series of QDs elongated along $[1\bar{1}0]$ with a lateral aspect ratio of 2 in \textit{absence} of the piezoelectric field [Fig.\ \ref{model} (a2)]. 

Fig.\ \ref{model} (b) shows the resulting FSS values for both series. The obtained trend for series 1 (piezoelectricity) is in qualitative accord with the experiment (Fig.\ \ref{data}). The importance of the piezoelectric potential for the discussion of the FSS stems from its strong dependence on QD height, which has already been shown earlier \cite{stier99}, independent of the chosen value of $e_{14}$. This explains why the FSS strongly increases with increasing QD size (decreasing exciton energy) [Fig.\ \ref{model}(b1)]. Quantitative agreement, however, could not be reached and deserves further investigation. A strong \textit{elongation} of the QDs in the $[1\bar{1}0]$ direction by contrast does not reproduce the experimentally observed trend [Fig.\ \ref{model}(b2)]. An elongation in the perpendicular $[110]$ direction (which is equivalent to reflecting the results on the x axis) also fails. It gives strongly negative FSS values for small QDs and slightly positive values for large QDs. Therefore elongation can be ruled out as the source of the observed FSS trend in the QDs investigated.

For small QDs the influence of piezoelectricity is weak and the FSS scatters around 0 $\mu$eV (Fig.\ \ref{data}). Then, second order effects like small random elongations of the QDs or interfacial symmetry lowering govern the FSS. In some QDs this leads to negative FSS values.

A considerable scatter of the observed FSS values of more than 100 $\mu$eV for a given energy is present in Fig.~\ref{data}. It is due to variations in QD geometry for a given QD height. A different base length for example leads to a different strain distribution inside the structure which in turn leads to different piezoelectric fields and hence to different FSS values. The variation in QD geometry for one QD height is visible through the observed FWHM of the subensemble peaks of around 30 meV in the PL spectrum \cite{heitz05}. Another source for the FSS scatter can be slight in-plane elongations which are more or less pronounced. 

To asses the influence of correlation on the FSS we compared calculations with a CI basis of 6 electron and 6 hole levels (C(6,6)) thus \emph{including} correlation effects to calculations with only 2 electron and 2 hole levels (C(2,2)) \emph{excluding} correlation. Qualitatively the results are the same, eliminating a decisive influence of correlation on our conclusions (Fig. \ref{model}).

In conclusion, for the first time a systematic change of the FSS with QD size is observed. We found a variation by one order of magnitude and observed a sign inversion of the FSS. Large values of up to 520 $\mu eV$ for large QDs were found. Complementing model calculations using \ekp\, theory and CI rule out elongation of the QDs as a source for the large FSS. To assess the influence of interfacial asymmetry in detail, atomistic theories need to be considered. Piezoelectricity can explain the size dependence but quantitatively fails to explain the huge variations observed.

Parts of this work were supported by Deutsche Forschungsgemeinschaft in the framework of SFB 296 and by the SANDiE Network of Excellence of the European Commission, Contract No. NMP4-CT-2004-500101. The electronic-structure calculations were performed on the IBM pSeries 690 supercomputer at HLRN within Project No. bep00014.

\end{document}